\documentclass[11pt]{article}
\usepackage{moriond,epsfig}

\def\be{\begin{equation}}
\def\ee{\end{equation}}
\def\bea{\begin{eqnarray}}
\def\eea{\end{eqnarray}}

\begin{document}
\vspace*{4cm}
\title{EMISSION AND STRUCTURE OF COMPACT FIREBALLS}

\author{ Amir Levinson }
           
\address{School of Physics and Astronomy, Tel Aviv University, Tel Aviv, Israel}

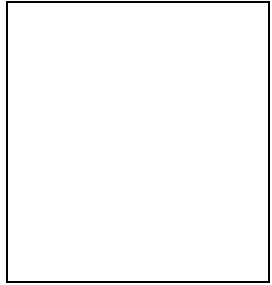
\maketitle\abstracts{The possibility that the peak of the spectral energy 
distribution (SED) of prompt GRB emission is produced on compact scales,
prior to the acceleration of the fireball to its terminal Lorentz factor
is considered.  It is argued that if dissipation on these scales is associated
with pair creation processes, then the observed clustering of SED peaks can 
be accounted for quite naturally by this mechanism.  Further dissipation and 
processing of pairs and gamma rays at larger radii, can give rise to a nonthermal
extension of the spectrum up to a rather high observed energy.  If the nonthermal
processes on these scales conserve the total flux of fireball quanta, then 
the SED peak, which is established on much smaller scales, will not be altered
significantly.  The asymptotic bulk Lorentz factor in this model should
exceed several hundred.
}

\section{Introduction}

In popular fireball models the dissipation of the fireball kinetic
energy is assumed to be accomplished through the generation of internal 
\cite{RM94,P99} and/or external \cite{MR93,CD99}
shocks.  In the so-called standard internal shock model (e.g, Rees \& 
Meszaros \cite{RM94}) it is conjectured 
that the fireball wind, expelled by a central source of dimension $\sim10^6$
cm, accelerates at small radii such that its Lorentz factor grows linearly 
with radius until the entire fireball energy is converted to kinetic energy.
Temporal fluctuations in the parameters of the ejected wind then lead to 
formation of internal shocks in the coasting phase (i.e., the region 
where the wind reaches its maximum Lorentz factor), at a distance of 
about $r_s\simeq \Gamma^2c\Delta t=3\times10^{11}(\Gamma/10^2)^2(\Delta t/
10^{-3} s)$ cm, where $\Gamma$ is the Lorentz factor of a disturbance in 
the injection frame, and $\Delta t$ is the variability time (i.e., the 
timescale over which the wind parameters fluctuate).  The radiation processes
commonly invoked are synchrotron and inverse Compton cooling of a nonthermal 
population of electrons (positrons) accelerated behind the (reverse and 
forward) shock fronts.  Depending upon the fireball parameters,
there may also be a contribution from photospheric emission and/or scattering
by turbulence \cite{T94,MR00}.

Several difficulties with the standard model have been noted in the literature:
Firstly, it has been argued that the efficiency with which the fireball 
energy is being converted into $\gamma$-rays is expected to be discouragingly 
small \cite{K99,SPM00}, which seems to be
inconsistent with the efficiency implied by afterglow observations.  
Secondly, the spectral energy distribution of GRBs peaks typically at 
around several hundreds keV, with relatively little scatter.  This clustering
of $\nu F_{\nu}$ peaks, if not due to observational selection effects
as now widely believed (but cf. Dermer et al. \cite{DBC99}), requires fine 
tuning if the emission near the peak 
originates from a region that moves with a large Lorentz factor,
as invoked in popular models \cite{EL00,MR00}.  Thirdly, the observed 
spectra below the SED peak appear to be steeper than predicted by the
synchrotron model \cite{Petal98,Fetal00}. 

These problems have motivated reconsideration of the standard internal 
shock model by a number of authors.  It has been shown that invoking 
very large variance in the Lorentz factors of different shells \cite{B00}
and/or multiple shell collisions \cite{GSW01}
may alleviate the efficiency problem.  Alternatively, radiative
viscosity mediated through the agency of some diffuse photons external
to the fireball can also lead to appreciable enhancement of the radiative 
efficiency of internal shocks \cite{L98}.  This may be particularly 
relevant in scenarios whereby the fireball is surrounded by a hot gas, 
as in the collapsor model \cite{W93} or the model discussed by Levinson 
\& Eichler \cite{LE93,LE00}.  Multiple shell collisions may also give rise to 
a correlation between the efficiency and peak energy, with the brightest 
bursts clustered in the observed energy range \cite{GSW01}, but only for 
certain choices of the Lorentz factor distribution of the ejected shells.

Below we consider an alternative scenario in which the SED peak of prompt
emission is produced on compact scales, prior to the acceleration of the
fireball to its terminal Lorentz factor. 

\section{The Compact Fireball Model}

Preferable peak energy and high efficiency can be naturally accounted for
if the emission near the peak originates from compact scales where the 
bulk Lorentz factor, $\Gamma_{0}$, is modest \cite{EL00}.  The requirements 
on the luminosity and peak energy yield the following relation between the 
Lorentz factor and radius of emission \cite{EL00}:
\begin{equation} 
r_{0}\simeq 10^{9.5}L_{52}^{1/2}\Gamma_{0}\epsilon_{300}^{-2}\ \ \ \
{\rm cm},
\end{equation}
where $\epsilon_{300}$ is the SED peak energy in units of 300 keV.

The basic picture envisioned is that at small radii, where the production
time of additional gamma-rays and pairs (e.g., thermalization time)
is sufficiently short, the fireball generates entropy as a result of 
dissipation, rather than being accelerating adiabatically 
as assumed in the standard model, so that 
the average photon energy measured in the observer frame decreases as the
fireball expands.  Once the average photon energy drops sufficiently 
below $m_ec^2$, the pair production process will taper off, at which point 
the SED peak is established.  The fireball will then start accelerating
to its terminal Lorentz factor.  The radius at
which the Thomson depth is roughly unity may be larger than the radius 
at which the SED peak is produced.  Nonetheless, the emitted spectrum should 
not look thermal (on both sides of the peak) if dissipation persists over 
a range of radii that encompass the Thomson photosphere (e.g., by a shock 
passing through the photosphere).  For example comptonization 
by energetic electrons (thermal or nonthermal) heated or accelerated 
locally, or by direct shock acceleration of photons \cite{BP81} can give 
rise to a power law extension of the spectrum up to a very high observed 
energy \cite{E94,EL00}.  If the dissipation leads only to redistribution 
of the fireball 
energy (that is, the net flux of pairs and gamma rays is conserved), then 
the average energy per quantum, as measured in the observer frame,
will remain constant (although the radial decrement of the average energy
in the fluid frame may be less steep than $1/r$).  The location of the 
SED peak in that case should not be altered significantly. 
(For a photon spectrum $\propto E^{-2}$, the break (minimum) energy 
is smaller than the average energy by a logarithmic factor.) 
    
Energization of the pair plasma at small scales
could come from internal shocks\footnote{It is often argued
that internal shocks cannot form in the region where the wind accelerates,
since in this region the Lorentz factors of different shells have the 
same radial variation ($\Gamma\propto r$) and, therefore, they cannot
catch up.  This is not necessary true if the wind profile deviates
from conical (as one might expect in the case of collimation), or if 
different shells satisfy different boundary conditions,
as, e.g., in the case wherein shell ejection occurs over a range of radii or
if the shells are oblique.} (e.g., refs \cite{RM94,E94,SP97}) but could 
also come from, say, collimation by 
surrounding baryonic material (a possibility that seems to be motivated
now by several considerations).  In the latter process, dissipation would 
proceed very efficiently when the average photon energy in the observer 
frame approached $m_ec^2$, for then large angle scattering by the walls 
of the collimating material of photons back into the jet, where they could 
appear in the local frame to be blueshifted to even higher energy, would 
lead to copious pair production.  

The location of the Thomson sphere is limited by the baryon load. 
The requirement that the photospheric radius, $r_{ph}$, should be smaller 
than the radius at which the entire burst energy is converted to 
kinetic energy of baryons implies a mass loss rate $\dot{M}<10^{28.5}
L_{52}^{7/8}\epsilon_{300}^{-1/2}$ gr s$^{-1}$,
$r_{ph}\le10^{12}L_{52}^{5/8}
\epsilon_{300}^{-3/2}$ cm, and $\Gamma_{ph}\le 300 L_{52}^{1/8}
\epsilon_{300}^{1/2}$, where $\Gamma_{ph}=\Gamma(r=r_{ph})$
\cite{EL00}.  If the photosphere were to occur at much smaller scales, then 
extremely small baryon load would be implied.  However, the photospheric
radius may be determined by nonthermal pair cascades and may be larger than
the limit set by baryonic contamination.

The maximum energy of accelerated electrons can be obtained by equating
the cooling and acceleration rates.  Assuming that the acceleration rate 
is a fraction $\eta$ of the electron 
gyrofrequency, and denoting by $\eta_B$ the magnetic field energy in units
of the equipartition value yields,
\begin{equation}
\epsilon_{max}\simeq1.7\eta_B^{1/4}\eta^{1/2}\Gamma^{3/2}r_{12}^{1/2}L_{52}^{-1/4}
\ \ \ {\rm GeV}.
\end{equation}
If only the beamed gamma-rays contribute to pair productions (that is, ignoring 
any contribution from external photons and/or large angle scattering by a 
surrounding matter), then the corresponding threshold energy reads: 
$\epsilon_{thrs}\simeq4\Gamma^2/\epsilon_{\gamma}$.  Taking 
$\epsilon_{\gamma}=\epsilon_{max}$ yields
\begin{equation}
\frac{\epsilon_{thrs}}{m_ec^2}\simeq10^{-3}\eta_B^{-1/4}\eta^{-1/2}
\Gamma^{1/2}r_{12}^{-1/2}L_{52}^{1/4},
\end{equation}
which is a fraction 
\begin{equation}
\frac{\epsilon_{thrs}}{\epsilon_{peak}}\simeq 3\times10^{-2}\eta_B^{-1/4}\eta^{-1/2}
\end{equation}
of the peak energy, where it has been assumed that $\Gamma$ increases linearly
with radius.  Consequently, efficient pair production would require 
$\eta_B^{1/4}\eta^{1/2}>10^{-1.5}$ or, alternatively, the presence of an 
additional source of unbeamed photons. 

In order to account for the observed afterglow emission, sufficient fraction of 
the burst energy should remain above the photosphere in the form of baryons, 
nonthermal pairs and/or magnetic fields.  In the absence of efficient generation 
of nonthermal pair cascades, as described above, a transition to a magnetically
dominated outflow just above the photosphere is required if the mass loss rate
$\dot{M}<<10^{28.5}$ gr s$^{-1}$ (see above), and is expected anyhow if the 
magnetic field energy beneath the photosphere is near its equipartition value. 
However, rapid pair production during the expansion 
of the fireball can convert a 
significant fraction of the energy into nonthermal pairs far enough out 
where the cooling and annihilation tims are long \cite{BL95}.
\section{Conclusions}
Dissipation on very compact scales, where the thermalization time or the 
timescale for generation of quanta by some other process is sufficiently 
short, can lead to an increase of the total flux of fireball quanta, and 
a consequent reduction of the observed energy per quantum.  This can give 
rise to a preferable peak energy below $m_ec^2$, particularly if effective
dissipation on these scales is associated with some pair creation processes;
as an example we proposed large angle scattering of gamma rays from a jetted
fireball back into the jet, by baryon rich wind surrounding the jet.  Additional
dissipation and processing of fireball quanta at larger radii, for instance by 
shocks moving along the ejecta, would result in
further evolution of the fireball spectrum and, ultimately, the formation of
nonthermal extension up to a rather high observed energy.  The 
location of the $\nu F_{\nu}$ peak might be affected by this additional dissipation,
but only slightly (roughly a logarithmic factor) if the flux of quanta is roughly
conserved.

\section*{Acknowledgment}
I thank D. Eichler for enlightening discussions.  This research was supported
by the Israel Science Foundation.


\section*{References}
\begin{thebibliography}{99}

\bibitem{RM94} M.J. Rees, \& P. M\'esz\'aros, {\em Astrophys. J.} {\bf 430}, 93 (1994)
\bibitem{P99} T. Piran, {\em Phys. Rep.} {\bf 314}, 575 (1999)
\bibitem{MR93} P. M\'esz\'aros, M.J. \& Rees, {\em Astrophys. J.} {\bf 405}, 278 (1993)
\bibitem{CD99} J. Chiang, \& C.D. Dermer, {\em Astrophys. J.} {\bf512}, 699 (1999)
\bibitem{T94} C. Thompson, {\em Mon. Not. R. Astron. Soc.} {\bf270} 480 (1994)
\bibitem{MR00} P. M\'esz\'aros, M.J. \& Rees, {\em Astrophys. J.} {\bf 530}, 292 (2000)\bibitem{K99} P. Kumar, {\em Astrophys. J.}, {\bf523}, L113 (1999)
\bibitem{SPM00} M. Spada, A. Panaitescu \& P. Meszaros, {\em Astrophys. J.} {\bf537}, 
824 (2000)
\bibitem{DBC99} C. Dermer, M. Bottcher \& J. Chiang, {\em Astrophys. J.} {\bf 515}, L49
(1999)
\bibitem{EL00} D. Eichler \& A. Levinson, {\em Astrophys. J.} {\bf529}, 146 (2000) 
\bibitem{Petal98}  R.D. Preece, et al., {\em Astrophys. J.} {\bf506}, L23 (1998)
\bibitem{Fetal00} F. Frontera, et al. {\em Astrophys. J. Supp.} {\bf127}, 59 (2000)
\bibitem{B00} A.M. Beloborodov, {\em Astrophys. J.} {\bf539}, 25 (2000)
\bibitem{GSW01} D. Guetta, M. Spada,\& E. Waxman, astro-ph/0011170
\bibitem{L98} A. Levinson {\em Astrophys. J.} {\bf507}, 145 (1998)
\bibitem{W93} S.E. Woosley, {\em Astrophys. J.} {\bf405}, 273 (1993)
\bibitem{LE93} A. Levinson D. \& Eichler {\em Astrophys. J.} {\bf418}, 386 (1993)
\bibitem{LE00} A. Levinson \& D. Eichler, {\em Phys. Rev. Lett} {\bf85}, 236 (2000)
\bibitem{BP81} R.D. Blandford \& D. Payne, {\em Mon. Not. R. Astron. Soc.} {\bf194}, 1041 (1981)
\bibitem{E94} D. Eichler, {\em Astrophys. J. Supp.} {\bf90}, 877 (1994) 
\bibitem{SP97} R. Sari, \& T. Piran {\em Mon. Not. R. Astron. Soc.} {\bf287}, 110 (1997)
\bibitem{BL95} R.D. Blandford, \& A. Levinson, {\em Astrophys. J.} {\bf441}, 79 (1995) 
 
\end{thebibliography}
\end{document}